# *Spitzer*'s Solar System Science Legacy: Studies of the Relics of Solar System Formation & Evolution. Part 1 - Comets, Centaurs, & Kuiper Belt Objects

(Nature Astronomy Manuscript NATASTRON19122828)

Carey Lisse[1], James Bauer[2], Dale Cruikshank[3], Josh Emery[4], Yanga Fernández[5], Estela Fernández-Valenzuela[6], Michael Kelley[2], Adam McKay[7], William Reach[8], Yvonne Pendleton[4], Noemi Pinilla-Alonso[5], John Stansberry[9], Mark Sykes[10], David Trilling[5], Diane Wooden[3], David Harker[11], Robert Gehrz[12], Charles Woodward[12]

*Pre-print Version 01-Jul-2020 revised for Nature Astronomy, ed. Paul Woods*

[1] Johns Hopkins University Applied Physics Laboratory, Laurel, MD 20723  carey.lisse@jhuapl.edu

[2] Department of Astronomy, University of Maryland College Park, College Park, MD  gerbsb@astro.umd.edu, dphamil@astro.umd.edu, msk@astro.umd.edu

[3] Space Science and Astrobiology Division, NASA Ames Research Center, Moffett Field, CA, USA 94035  dale.p.cruikshank@nasa.gov, yvonne.pendleton@nasa.gov, diane.wooden@nasa.gov

[4] Department of Astronomy and Planetary Sciences, Northern Arizona University, Flagstaff, AZ 8601  joshua.emery@nau.edu, David.Trilling@nau.edu

[5] Department of Physics & Florida Space Institute, University of Central Florida, Orlando, FL 32816  Yanga.Fernandez@ucf.edu, npinilla@ucf.edu

[6] Florida Space Institute, University of Central Florida, Orlando, FL, 32826  sestela@ucf.edu

[7] NASA Goddard Spaceflight Center, Greenbelt, MD 20771 and American University  adam.mckay@nasa.gov

[8] Stratospheric Observatory for Infrared Astronomy, Universities Space Research Association, NASA Ames Research Center, Moffett Field, CA 94035  wreach@sofia.usra.edu

[9] Space Telescope Science Institute, 3700 San Martin Dr. Baltimore, MD 21218  jstans@stsci.edu

[10] Planetary Science Institute, Tucson, AZ 85719  sykes@psi.edu

[11] University of California, San Diego, Center for Astrophys. & Space Sci., La Jolla, CA 92093  dharker@ucsd.edu

[12] Minnesota Institute for Astrophysics, University of Minnesota, 116 Church Street, S. E., Minneapolis, MN 55455  gehrz@astro.umn.edu, chickw024@gmail.com

22 Pages, 9 Figures, 0 Tables

Proposed Running Title: **"*Spitzer*'s Solar System Science Legacy: Studies of the Relics of Solar System Formation & Evolution. Part 1 - Comets, Centaurs, & Kuiper Belt Objects"**

Please address all future correspondence, reviews, proofs, etc. to:

Dr. Carey M. Lisse

Planetary Exploration Group, Space Exploration Sector

Johns Hopkins University, Applied Physics Laboratory

SES/SRE, Building 200, E206

11100 Johns Hopkins Rd

Laurel, MD 20723

240-228-0535 (office) / 240-228-8939 (fax)

Carey.Lisse@jhuapl.edu

# Abstract

In its 16 years of scientific measurements, the *Spitzer* Space Telescope performed a number of ground breaking and key infrared measurements of Solar System objects near and far. Targets ranged from the smallest planetesimals to the giant planets, and have helped us reform our understanding of these objects while also laying the groundwork for future infrared space-based observations like those to be undertaken by the James Webb Space Telescope in the 2020s. In this first Paper, we describe how the *Spitzer* Space Telescope advanced our knowledge of Solar System formation and evolution via observations of small outer Solar System planetesimals, i.e., Comets, Centaurs, and Kuiper Belt Objects (KBOs). Relics from the early formation era of our Solar System, these objects hold important information about the processes that created them. The key *Spitzer* observations can be grouped into 3 broad classes: characterization of new Solar System objects (comets D/ISON 2012 S1, C/2016 R2, 1I/`Oumuamua); large population surveys of known object sizes (comets, Centaurs, and KBOs); and compositional studies via spectral measurements of body surfaces and emitted materials (comets, Centaurs, and KBOs).

# I.     Introduction

NASA's *Spitzer* Space Telescope (hereafter *Spitzer*, Werner *et al.* 2004, Gehrz *et al.* 2007) produced an amazing array of science results during its 16-year mission in trailing Earth orbit. Highly sensitive at 3.6 to 160 um infrared wavelengths and able to observe objects in our Solar System from vantage points different than Earth's, it provided us with unique views of small Solar System bodies and outer Solar System objects. In this paper we review some of *Spitzer*'s greatest planetary science findings, from ~1 au to the edge of our Solar System.

It has been said that *Spitzer*'s scientific forte was "the old, the cold, and the dusty" in our universe (Werner & Eisenhardt 2019). In its 5.5 years of 3-160 um cryogenic operations and 10.8 years of subsequent warm era operations observing the Solar System from Aug 2003 to Jan 2020, *Spitzer*'s observing capabilities were mainly sensitive to small relic solids in interplanetary space, as its infrared sensors were so sensitive that it saturated on any large (i.e., radius > 300 km) moon or planet-like body inside the orbit of Uranus. But in order to keep sunlight from warming the spacecraft, *Spitzer* was also unable to look at objects closer to the Sun than 82.5º elongation, i.e., the inner system asteroids. Thus in the Solar System, we need to say that *Spitzer*'s scientific forte was "the old, the cold, the dusty, and the small or faraway".

That being said, *Spitzer* produced seminal science on a plethora of Solar System objects: the Zodiacal Cloud (Zody) and Near Earth Asteroids (NEAs) at ~1 au; main belt asteroids (MBAs) at 2-5 au; comets from 1 – 30 AU; Jovian Trojans at ~ 5.2 AU; Saturn's moon Phoebe at 10 AU; Centaurs from 5 to 30 AU; Uranus at 18 AU; Neptune at 30 AU; and Kuiper Belt Objects (KBOs) from 30 – 50 AU. The small bodies that *Spitzer* has studied are all (or are closely related to) the relic planetesimals left over from the era of planetary formation. The discovery by the Kepler/K2 mission that Neptune-sized planets are the most frequently-occurring type of exoplanet (Fulton *et al.* 2017) means that *Spitzer* studies of our own ice giants, Uranus and Neptune, are of additional, broader importance to understanding planetary system formation. We thus ask the reader to keep the current Solar System formation paradigm in mind (DeMeo & Carry 2014, Fig. 1) when reading further about the zoo of objects that *Spitzer* observed orbiting around a 4.56 Gyr old main sequence G2V star located in the disk ~ 2/3 of the way out from the core of an ~12 Gyr old spiral galaxy.

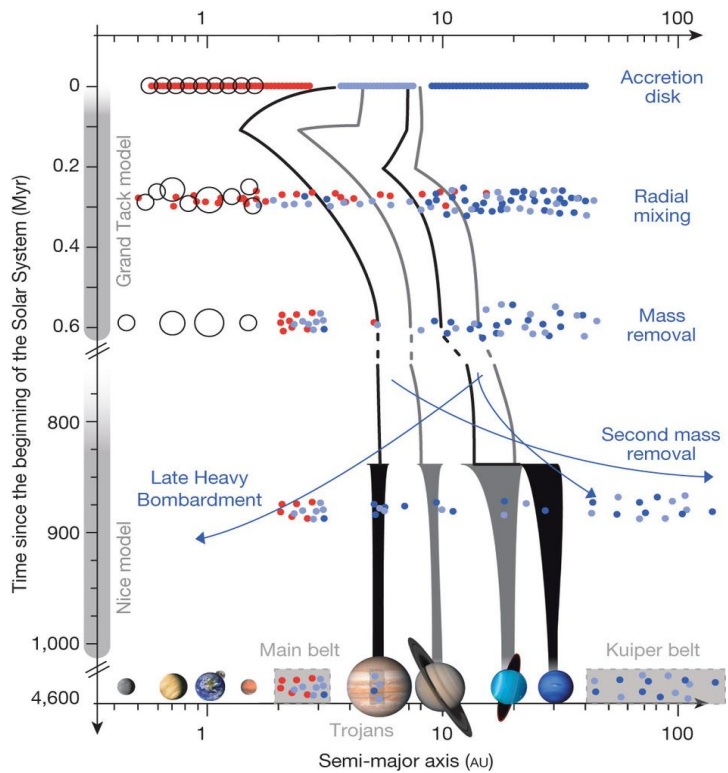

**Figure 1 - The current paradigm of planetary system formation** invokes the condensation of rocky refractory materials and condensible ices in the thick proto-solar disk, surrounded by a gaseous envelope. Eventually the solid grains settle into a thin disk, whereupon aggregation into larger bodies begins. Dust grains grow to become pebbles, then boulders, then kilometer-sized and hundred-kilometer-sized planetesimals - comets and asteroids - the building blocks of solid planets. Subsequent planet formation has led to the clearing of the Sun's disk inside about ~30 AU. Many planetesimals in the inner Solar System were incorporated into the present-day terrestrial planets, but many persist as 1- to 1000-km asteroids. Planetesimals that formed in the region of the gas giants were either incorporated into the planets or rapidly ejected. Most escaped the Solar System entirely, but a large population was captured by stellar and galactic perturbations into the Oort cloud. Oort comets may be gravitationally scattered back into the inner Solar System to appear as dynamically new long-period comets, some of which are eventually captured into Halley-family short-period orbits. In the cool outer disk (beyond Jupiter's orbit), the planetesimals incorporated frozen volatiles as well as refractory material. The trans-Neptunian region is occupied by numerous bodies with sizes up to several thousand km (Pluto and other dwarf planets). Collisions in this population generate fragments, some of which are scattered inward by dynamical chaos to become Centaurs in the giant planet region, and then Jupiter-family short-period comets in the inner Solar System. Comets that lose all their volatiles from prolonged solar heating appear as asteroids having unusual, comet-like orbits (after DeMeo & Carry 2014).

In order to better focus *Spitzer*'s Solar System Legacy Science results, this paper has been published in two sections, Part 1 and Part 2. In the next sections of Part 1 presented below, we briefly describe *Spitzer*'s findings for the inter-related small planetesimals of the outer Solar System (currently thought to have formed from their own outer proto-planetary disk reservoir once Jupiter formed) and provide the reader with key references for further reading & research. Our review here follows in the footsteps of earlier reviews of cold-phase *Spitzer* Solar System observations (e.g., Cruikshank 2005, Gehrz *et al.* 2006, Fernandez *et al.* 2006, Werner *et al.* 2006) but seeks to include the multitude of discoveries from *Spitzer*'s lengthy warm era.

## II. Comets

**IIa. Comet Nuclei.** Comets are small icy relic planetesimal bodies left over from the original condensation of Interstellar Medium (ISM) gas and dust in the protoplanetary disk. As such, they are objects containing important clues to planetary formation in the early Solar System, requiring detailed characterization. In 2006 – 2007 *Spitzer* photometrically measured the thermal emission from 89 Jupiter-family comets (JFCs) under the cold-era Survey of the Ensemble Physical Properties of Cometary Nuclei (SEPPCoN) program of Fernandez *et al.* (2013). The survey at the time provided the largest compilation of radiometrically-derived physical properties of nuclei. Virtually all of the comets were observed while over 3 AU from the Sun in order to minimize the chance of any coma being present, and indeed 54 of the comets appeared bare. Interestingly, 35 comets displayed discernible dust in a surrounding coma, and 21 of those were actually active at the time of observation despite being so far from the Sun (Kelley *et al.* 2013; Fig. 2).

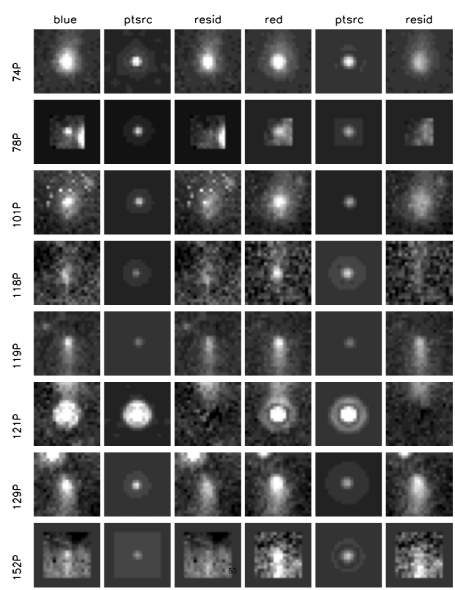

**Figure 2 - Spitzer "Blue" (16 um IRS Peakup Camera) and Spitzer "Red" (22 um Peakup Camera) deep imaging of 8 'distant' (= 3-6 AU from the Earth & Spitzer) comets in the thermal IR.** Figure is from Fernandez et al. (2013). Heat radiation from both the comet's nucleus and surrounding dust is detected. The best fit unresolved nucleus point source ('ptsrc') is separated out from the extended dust emission using the coma modeling technique of Lisse *et al.* (1999) and the nucleus size calculated using the NEATM model of Harris 1998.

Major findings from the survey included: (a) Effective radii measured for all 89 comets. The measured thermal emission from the nuclei was consistent with negligible thermal inertia. (b) The observed JFC cumulative size distribution (CSD; Figure 3) – represented as a function that shows the number of nuclei with a radius larger than a given value, and so is monotonically decreasing (cf. Lamy *et al.* 2004) – has a power-law slope of around -1.9, similar to that derived by other groups from visible-wavelength observations, implying that the geometric albedo (hereafter 'albedo') of cometary nuclei does not strongly track with radius. (c) The CSD is consistent with an intrinsic distribution that does not have many sub-kilometer objects (Meech *et al.* 2004), similar to the KBO impactor CSD deduced from studying Pluto system craters (Singer *et al.* 2019, Lisse *et al.* 2019), which suggests that primordial icy planetesimals "formed large". (d) There was a significant perihelion dependence on the distant activity, with none of the JFCs that come within 1.8 AU of the Sun found to be active in the outer parts of their orbits.

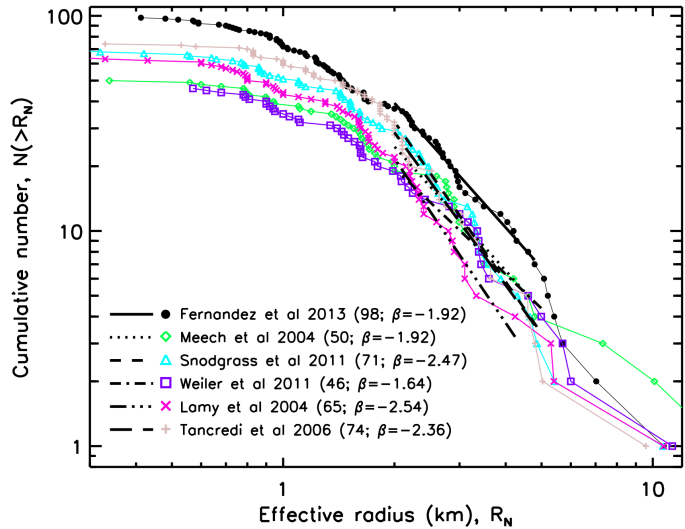

Figure 3 – **Cumulative size distribution of the JF comets as derived from Spitzer observations** (solid line and filled circles) in comparison to other reported measurements using visible-wavelength observations that must assume an albedo (other curves/symbols). The plot shows the number of cometary nuclei with a radius larger than a given value. In the legend, β refers to the power-law slope of the line. This dependence is very similar to that found for the KBOs from cratering studies of Pluto & Charon (Singer *et al.* 2019). Figure is adapted from Fernandez et al. (2013).

*Spitzer* also provided some of the first thermal measurements of the nuclei of the Activated Asteroids, also known as the Main Belt Comets. This population was first noted in the late 1990s with the discovery of extended emission from the ostensible asteroid (7968) Elst-Pizarro (now also known as 133P/Elst-Pizarro; Hsieh & Jewitt 2006). While there are still less than two dozen such objects known, and since not all of these objects' extended emission is due to cometary activity (Jewitt 2012), a size distribution is not yet feasible. However *Spitzer* has helped to open the door to characterizing this unexpected population that is bringing new ideas to the preservation of volatiles in environments there were heretofore thought to be rare.

Finally, *Spitzer* provided us insight into remnant activity of low-Tisserand asteroids that are actually dormant comet nuclei. For example, cometary activity was seen for the first time from (3552) Don Quixote (Mommert *et al.* 2014, 2020) in the form of a gas coma in the *Spitzer* Infrared Array Camer (IRAC; Fazio et al. 2004) Channel 2, suggesting hypervolatiles remain in this presumably highly-evolved object. We note that this was a rare instance of cometary activity coming from a dormant or near-dead comet that manifested itself as gas instead of as dust. The only other possible instance of this happening is the reported 1949 activity of comet 107P/Wilson-Harrington (Fernandez *et al.* 1997).

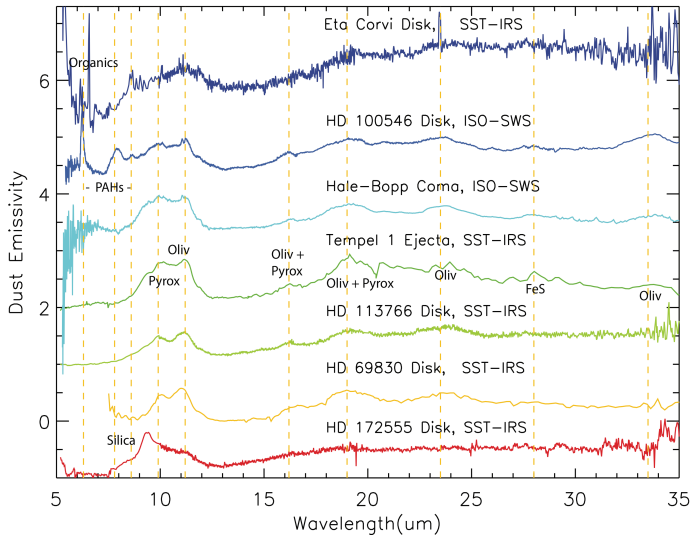

**Figure 4a - Spitzer/IRS emissivity spectra of dust from SP comet 9P/Tempel 1 (Lisse+ 2006, 2007a),** compared to that of **LP comet C/Hale-Bopp 1995 O1 (ISO, Crovisier+ 1997)** and of 5 nearby bright debris disk systems taken by Spitzer (~1.4 Gyr old F2V η Corvi; ~10 Myr old A0Ve system HD100546, Lisse+ 2007a; ~12 Myr F2V HD113766A, Lisse+ 2008; 6-10 Gyr old HD69830, Lisse+ 2007b; and ~20 Myr HD172555, Lisse+ 2009). Common spectral features due to carbonaceous materials at 6-8 um, ferromagnesian silicates at 9 – 22 um, and ferromagnesian sulfides at 25 – 28 um are seen in the first 4 spectra, making the case for exocomets dominating the young Jovian planet building HD100546 and mature LHB η Corvi disks. By contrast, young HD113766A appears to building a rocky terrestrial planet out of hydrated exoasteroidal material, very mature HD69830 disk is dominated by debris from an exoasteroid-exoasteroid collision (see Part 2, Trilling+ 2020). HD172555 is full of unusual high temperature altered silica formed in a late-stage rocky planet giant impact (see the Spitzer Circumstellar Legacy Science review of Chen+ 2020).

**IIb. Comet Composition.** *Spitzer* Infrared Spectrograph (IRS; Houck *et al.* 2004) spectral observations of comets whose interior materials were released into the coma from quiescent activity (Kelley *et al.* 2006; Woodward *et al.* 2006, 2011), massive outflows (C/1995 O1 (Hale-Bopp), Crovisier *et al.* 1997, Wooden *et al.* 1999), impact (9P/Tempel 1, Lisse *et al.* 2007), fragmentation (73P/Schwassmann-Wachmann 3, Reach *et al.* 2009, Sitko *et al.* 2011), or explosive eruption (17P/Holmes, Reach *et al.* 2010), showed a rich variety of ferromagnesian silicates, sulfides, amorphous carbon, carbonates, and PAH components, similar to those found by the ROSETTA spacecraft in comet 67P/C-G (Levasseur-Regourd 2018, Mannel *et al.* 2019) and orbiting around other stars in exodisk systems observed by *Spitzer* (Figure 4a & 4b). This group of materials is close to what one might expect as a result of a slowly cooling mix of protoplanetary disk material of solar composition that is well mixed and heated above 1000 K by occasional close passages by the Sun (Lewis & Prinn 1997), and matches well the materials found in the Stardust mission sample returns (Westphal *et al.* 2017).

The silicate spectral features found in *Spitzer* mid-infrared spectra of the nuclei of comets 49P/Arend-Rigaux and 10P/Tempel 2 matched those seen in D-type Jovian Trojans (Kelley *et al.* 2017, see also Part 2, Fig. 3, Section III Trojans). Cometary nuclei were known to have D-type reflectance spectra (red, but otherwise featureless). However, the commonality in mid-infrared silicate features suggests a possible genetic link between cometary nuclei and D-type asteroids.

The fragmentation of comet 73P/Schwassmann-Wachmann 3 (Sitko *et al.* 2011) into multiple daughter fragments revealed that the interior bulk composition of the original nucleus appeared homogeneous, as the SEDs from fragments B and C had similar grain composition as deduced from thermal modeling (Fig. 4b). In addition, the study of 73P/SW3 also demonstrated that comets could have a large amorphous carbon content. Amorphous carbon is a prime candidate for the dark spectrally featureless organic matter on a number of *Spitzer* comets (Lisse *et al.* 2006, 2007; Reach *et al.* 2010; Sitko *et al.* 2011; Wooden *et al.* 2017) and thermal modeling of other *Spitzer*/IRS comet fluxes (e.g., Woodward *et al.* 2011, Harker *et al.* 2018) clearly demonstrates that comets contain significant amorphous carbon content. This is consistent with *in situ* characterization of the 67P dust composition by Rosetta COSIMA (Bardyn *et al.* 2017), which shows that 50% of the elemental carbon in the comet is in higher molecular weight organics most akin to carbonaceous chondrite insoluble organic matter. The carbon content in comets is now viewed as a new diagnostic for compositional gradients in proto-solar disks that are possibly linked to the formation of terrestrial-like planetary bodies (Bergin *et al.* 2015).

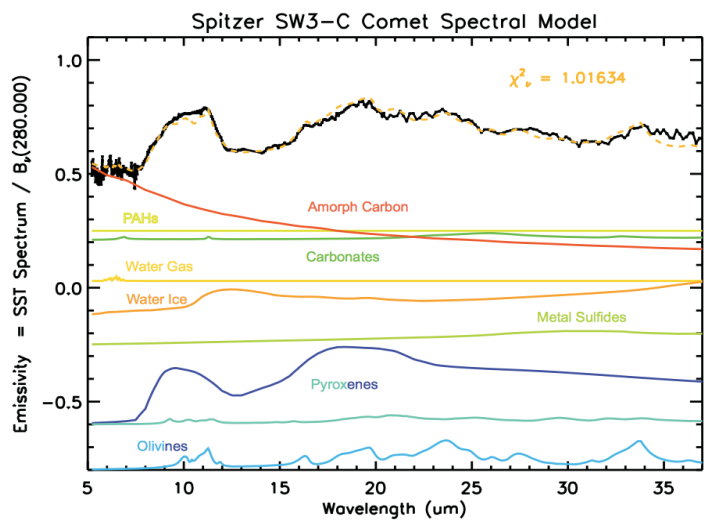

**Figure 4b - Spitzer/IRS spectrum of 73P/SW3 fragment C (black line)** and the corresponding thermal model decomposition (orange dashed line), showing the wealth of carbonaceous, aqueous, ferromagnesian silicate, and ferromagnesian sulfide materials in this comet's coma. Similar to fragment 73P/SW3 B, the pyroxene is highly amorphous, and the comets contain very large amounts of primitive refractory carbon, suggesting 73P formed in a very unprocessed part of the solar nebula. Adopted from Sitko *et al.* (2011).

*Spitzer* was one of the first facilities able to give us significant information on the abundance of $CO_2$ in comets thanks to the happy situation of the $CO_2$ ν3 asymmetric stretch band near 4.27 μm falling within the bandpass of IRAC's Channel 2. Unfortunately, a CO (1,0) band near 4.67 μm also falls within the bandpass, so the interpretation of a coma that appears in Channel 2 imaging can be sometimes uncertain. However the fluorescence efficiency of $CO_2$ in this band is about an order of magnitude higher than that of the CO band, so for comets where the CO and $CO_2$ abundances are comparable, the gas coma can be safely interpreted to first order as coming from $CO_2$. This is significant since there is basically no way to observe cometary $CO_2$ from the ground. An early *Spitzer* survey of cometary $CO_2$ using this methodology was performed by Reach *et al.* (2013) who found many comets with a $CO_2$ gas coma, as well as a correlation between the relative strength of this $CO/CO_2$ gas coma and the "depleted"/"normal" classification in C2/C3 carbon-chain daughter species found by A'Hearn *et al.* (1995). Reach *et al.* (2013) also discovered $CO/CO_2$ gas coma morphology – in the form of gas jets and spirals – that was not connected to features in the dust coma.

## IIc.     Unique Comets & Interstellar Comets

*Spitzer* also observed a number of unique, paradigm changing comets. It observed and characterized 4 comet nuclei "spacecraft mission" targets: 9P/Tempel 1 in 2004 – 2005 in support of the Deep Impact excavation mission (A'Hearn *et al.* 2005, Lisse *et al.* 2005, 2006); 67P/C-G in 2004-2007 for the ROSETTA rendezvous mission (Kelley *et al.* 2008, 2009; Agarwal *et al.* 2010); 8P/Tuttle in 2008, in support of the Arecibo Radio Observatory's "mission" to that close-flyby comet (Lamy *et al.* 2008, Harmon *et al.* 2010, Groussin *et al.* 2019); and 103P/Hartley in 2009 for the EPOXI flyby mission (A'Hearn et al. 2011s, Lisse *et al.* 2009). In 2013, it observed C/2012 S1 (ISON) before it passed close by Mars, detecting the comet's increasing abundant gas and $CO/CO_2$ dust emission (Lisse *et al.* 2013, Meech *et al.* 2013), and then again after it exploded near the Sun, leaving behind only a fan-shaped dusty debris trail headed back out from the Sun towards the Earth (Knight *et al.* 2013, priv. commun). In 2018 simultaneous *Spitzer* and ground-based observations helped determine the $H_2O$, $CO_2$, and CO gas production rates from comet C/2016 R2 (PanSTARRS) (Cochran & McKay 2018, Biver *et al.* 2018, Wierzchos & Womack 2018, McKay *et al.* 2019, Opitom *et al.* 2019). By comparing the *Spitzer* data to observations of CO and $H_2O$ obtained with other facilities, McKay *et al.* (2019) determined a small $CO_2/CO$ ratio of 0.18 and a huge $CO_2/H_2O$ ratio of ~3230, consistent with a uniquely cold (~30 K) cometary object dominated by hypervolatile CO ice sublimation, and the best candidate we now have for a piece of primordial KBO material.

Perhaps the "Solar System" object originating from the farthest out region ever observed by *Spitzer* was 1I/`Oumuamua (hereafter 1I), discovered in October 2017 by Williams *et al.* 2017. Its orbit was determined to be hyperbolic, which can happen for barely bound Oort Cloud comets that get a gravitational perturbation from one of the giant planets as it passes through the inner solar system. However, 1I's eccentricity = 1.2 was so large that it could not be produced by any known solar system perturbation, and it was moving so fast that it left the solar system in 2018 – 2019 with an excess relative speed $v_\infty$ = 26 km/sec. It must therefore have been unbound to the solar system and have originated in another planetary system. The object was not detected by *Spitzer* in November 2017, which gives an upper limit of 0.3 µJy at 4.5 um (3σ) Trilling *et al.* (2018). Using the NEATM model and optical photometry of 1I along with the *Spitzer* upper limits, Trilling *et al.* found that 1I likely had thermal emission like a comet nucleus, but an albedo higher than cometary. This high albedo corresponds to a small diameter, but also implies the exposure or creation, from outgassing, of a fresh, icy, bright surface due to thermal reactivation during 'Oumuamua's close perihelion passage ($r_{h,perigee}$ = 0.25 au) in 2017 September. *Spitzer*'s very low upper limits on CO and $CO_2$ outgassing from 1I argue that this bright surface is likely composed of water ice.

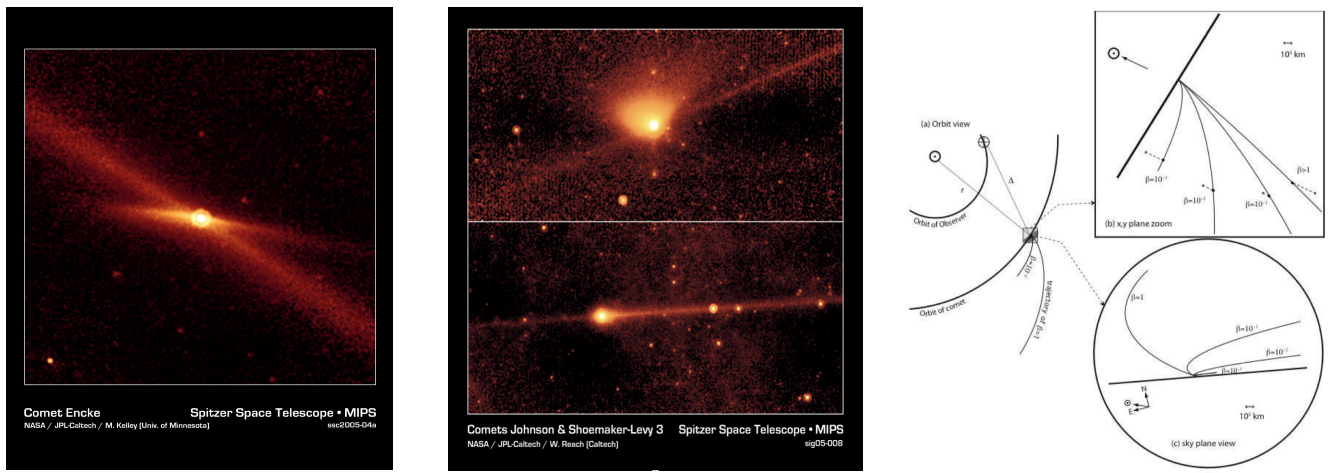

**Figure 5 - Examples of Spitzer observations of large-particle-dominated trails** include 2P/Encke (***left***) and 48P/Johnson and 129P/Shoemaker-Levy 3 (***middle***), for which the viewing geometry allowed straightforward separation of particles of different size. The contrail-like structure of trails arise from their large particles having associated low βs (insensitive to radiation pressure), resulting in their remaining in proximity to the parent comet orbit, while higher β -particles associated with tails and coma extend into the plane of the comet orbit away from the Sun and the comet orbit. To distinguish them there must be some degree of angular separation above or below the orbital plane (***right***).

**IId.    Comet Trails.** The discovery of comet debris trails in the 1983 IRAS sky survey (Davies *et al.* 1984, Sykes *et al.* 1986) revealed significant emissions of large (mm-cm) particles from cometary nuclei, that such particles were the dominant mechanism by which comets lose mass, and that comets were far rockier than previously thought, having average dust-to-gas mass ratios of ~3 instead of the canonical 0.1-1 (Sykes and Walker, 1992). Although IRAS only detected eight trails, it was hypothesized that trails should be common to all short-period comets (Sykes and Walker, 1992) and that comet trails are the major supply source of the interplanetary dust cloud (Nesvorny *et al.* 2010).

*Spitzer* afforded the first opportunity to test these results. Reach *et al.* (2007) observed 34 comets using the Multi-Band Imaging Photometer (MIPS) at 24 μm along at least $10^6$ km of each comet's orbit. Debris trails due to mm-sized or larger particles were found along the orbits of 27 comets (Fig. 5); 4 comets had small-particle dust tails and a viewing geometry that made debris trails impossible to distinguish; and only 3 had no debris trail despite favorable observing conditions. This extended the known debris trails associated with Jupiter-family comets to 30. The detection rate by *Spitzer* was > 80%, confirming that debris trails are a generic feature of short-period comets. By comparison to orbital calculations for particles of a range of sizes ejected over 2 yr prior to observation, particles dominating the surface area of 4 debris trails were typically mm-sized while the remainder of the debris trails required larger particles. The lower-limit masses of the debris trails were typically $10^{11}$ g over the range of orbit observed, and the median mass loss rate was determined to be 2 kg/s. The mass-loss rate in trail particles was comparable to that inferred from OH production rates and larger than that inferred from visible-light scattering in comae.

Some correlation was found by Reach *et al.* (2007) between the Tisserand invariant $T_{Jup}$ (a dynamical quantity approximately conserved in the 3-body Sun-Jupiter-comet system) for Jovian perturbations for different comets orbits with trails, and mass loss rates (dM/dt). Comets with little interaction with Jupiter (asteroidal or Encke-type orbits, $T_{Jup}$ > 2.94) had a median dM/dt = 3.4 kg s$^{-1}$, while those with $T_{Jup}$ < 2.94 had a median dM/dt = 0.8 kg s$^{-1}$. These trends suggested that comets spending more time closer to the Sun may produce more debris, and comets dynamically decoupled from Jupiter may retain debris longer.

Outside of the Reach et al. (2007) survey, dust trails were detected at several comets with *Spitzer*, e.g., 2P/Encke, 29P/Schwassmann-Wachmann 1, 103P/Hartley 2, 144P/Kushida, 260P/McNaught (Stansberry et al. 2004, Nesvorný et al. 2006, Lisse et al. 2009, Reach et al. 2010, Vaubaillon & Reach 2010, Kelley et al. 2013). Perhaps the most spectacular cometary nucleus – cometary trail complex observed to date was the 73P/Schwassmann-Wachmann 3 (hereafter 73P/SW-3) complex imaged by Reach *et al.* (2009, 2010) in 2006 using *Spitzer*/MIPS at 24 um. Over 60 fragments are painted along the orbit of the main parent nucleus that is easily delineated across the sky by the comet's trail, and it can easily be seen that the trail is a continuation of the mass shedding process for this comet. Deep MIPS observations were also performed for the dust trail of comet 67P, important as pre-encounter information for the Rosetta mission as well as intrinsic interest in its narrow trail and close approaches to Earth (Kelley *et al.* 2008, 2009). Agarwal *et al.* (2010) used this data + dynamical models to separate emission from different-sized particles, concluding that the cross-section of cometary mass loss is primarily contributed by particles in the 60 to 600 micron size range.

Since the pioneering *Spitzer* Comet Trail research, a re-analysis of the COBE/DIRBE 1990 infrared all-sky photometric survey by Arendt (2014) has turned up pre-*Spitzer* recoveries of the 2P/Encke and 73P/SW-3 trails, demonstrating their stability over multiple orbits, while new trails were discovered for long-period comet 1P/Halley and potentially hazardous comet-asteroid transition object 3200 Phaethon. [The detection 1P and 2P's trails solves a major mystery from Lisse et al. 1998's COBE/DIRBE analysis, as these two comets had long been predicted (e.g. by Kresak & Kresakova 1987) to be the major contributors of new bound dust to the local interplanetary dust environment.] Trails have now been observed in visible scattered light as well, first serendipitously by Rabinowitz and Scotti (1991) for 4P/Faye and then in more focused studies for 22P/Kopff by Ishiguro et al. (2002), 81P/Wild 2 (Ishiguro et al., 2003), and 67P/Churyumov-Gerasimeko (Ishiguro 2008).

## III.     Centaurs

The evolutionary link between inner system Jupiter Family Comets (JFCs) and their source population Kuiper Belt Objects (KBOs; Section IV) manifests itself in the Centaur population, small planetesimals on transient orbits among the giant planets (cf. Lamy & Toth 2009). The precise dynamical parameters that categorize small bodies as Centaurs has previously been in dispute (cf. Gladman *et al.* 2008, Jewitt 2009, Levison & Duncan 1997). However, there is general agreement that bodies which cross the orbits of the giant planets, i.e., with semi-major axis values between 5.2 and 30 au (Jewitt and Kalas 1998), are on dynamically short-duration orbits that evolve quickly, on the order of a few tens of millions of years. This dynamical behavior enables a significant observational opportunity to study these bodies.

Not completely primitive and not as evolved as JFCs, Centaurs are undergoing the processes of physical change between the objects in the "deep-freeze" relatively static state of KBOs and the active and thermally altered state of cometary bodies, i.e., volatile bearing bodies that undergo routine insolation of relatively large magnitudes. Surface processing, both chemical and physical, mass loss in the form of dust and gas, and the change in size of the JFCs over time are likely initiated while the body is in its Centaur stage. The observations by *Spitzer* have provided key insights to the study of the Centaur population, reflecting instances of the transformation of KBOs into comets in mid-stride.

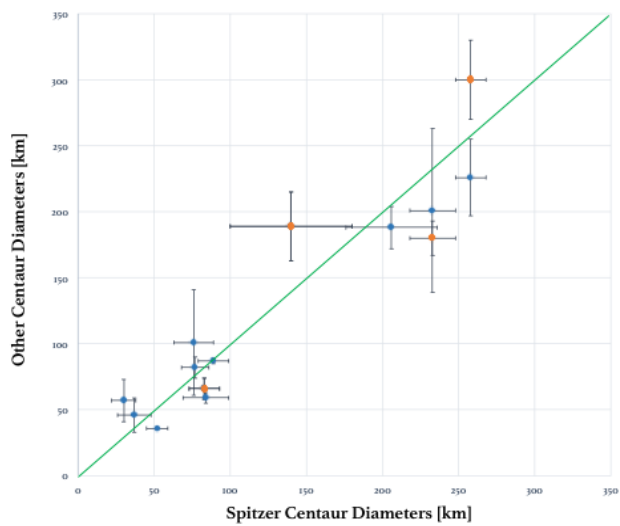

**Figure 6 - Spitzer-derived Centaur diameters as compared to those derived from other measurement methods.** The values show good overall agreement; those that overlap the green line match within the stated uncertainties. The blue circles shown are for values from Bauer *et al.* (2013) and the orange from Fernandez *et al.* (2002).

**IIIa. Centaur Size and Albedo Surveys.** *Spitzer* led the way in obtaining statistically meaningful Centaur sizes by expanding the total of known Centaur diameters from 4 to 20 (c.f., Fernandez *et al.* 2002) with thermal photometric observations (Stansberry *et al.* 2008). Figure 6 summarizes these diameter measurements and shows that the initial radiometrically derived diameters compared well with the preceding measurements of Pholus (Davies *et al.* 1993), Chiron (Bus *et al.* 1996), Chariklo (Jewitt and Kalas 1998), and Asbolus (Fernandez *et al.* 2002).

The sample of Centaur thermal measurements obtained by *Spitzer* suggested that:
- Mean Centaur albedos are a few percent, similar to comets and in agreement with later results from the key program of the Herschel Space observatory, "TNOs Are Cool" (Duffard *et al.* 2014, Müller *et al.* 2020)

- The color groupings of the centaur populations (Peixhino *et al.* 2003) correlate with surface reflectance; gray centaurs have on average darker albedos than those exhibited by the redder centaurs (Tegler *et al.* 2008).
- The average Centaur beaming parameters, which can be driven by a combination of surface characteristics including thermal inertial and surface roughness, are close to unity and are also similar to those of comets (Fernandez *et al.* 2013).

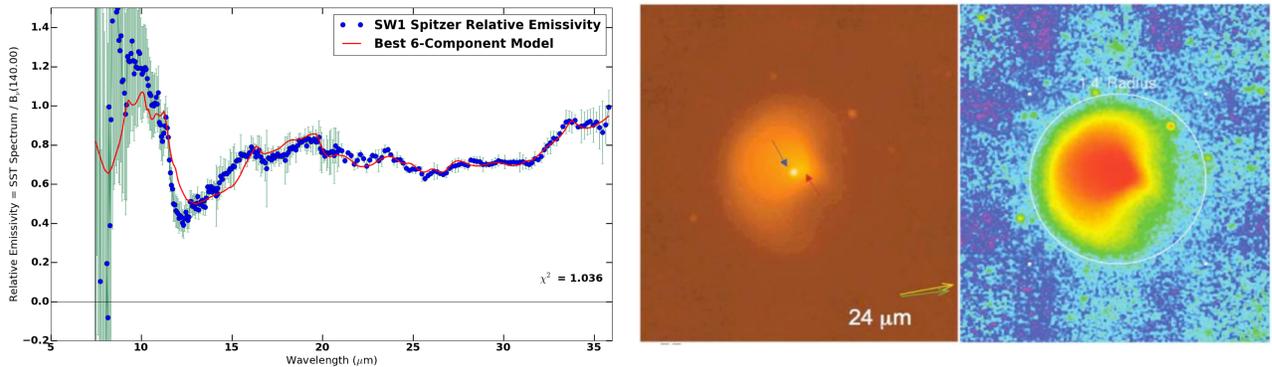

**Figure 7 - Spitzer-thermal measurements of two active Centaurs.** *(left)* IRS spectrum of 29P/Schwassmann-Wachmann 1 (Stansberry et al. 2004) divided by the LTE blackbody function to show emissivity; figure is from Schambeau et al. (2015). A six-component composition model (red line) has been fit to the emissivity spectrum; the best-fit includes amorphous carbon, water ice, amorphous olivine, forsterite, diopside, and metal sulfides, *(right)* The 2006 outburst of the active Centaur 174P/(60558) Echeclus from Bauer *et al.* (2008), showing significant dust signal at 24 and 70 microns.

The NEOWISE (Near-Earth Object Wide-field Infrared Survey Explorer) survey (Mainzer *et al.* 2011) and the Herschel space observatory (Muller *et al.* 2020) provided Centaur detections that augmented the results obtained by *Spitzer* (Bauer *et al.* 2013, Duffard *et al.* 2014). The combined sample of *Spitzer* and NEOWISE thermal measurements confirmed the above conclusions, and also revealed a Centaur size distribution with slope parameter similar to JFCs and KBOs, affirming a genetic relation between the three populations.

**IIIb. Active Centaurs.** The first Centaur discovered was identified as a comet. The Centaur 29P/Schwassmann-Wachmann 1, which appears to have continuous activity at heliocentric distances ~6 AU, has been measured to produce large quantities of CO (e.g. Senay & Jewitt 1994, Paganini *et al.* 2013, Wierzchos and Womack 2020), ~$10^{28}$ molecules per second. The second Centaur discovery, 2060 Chiron (Kowal 1977), was also found to be active (cf. West 1991), and out to distances beyond 12 au (Meech & Belton 1990), also attributed to CO production (Womack *et al.* 2017). Approximately 8% of the known Centaurs have had confirmed activity, indicating a large fraction of the Centaur population contains significant volatile abundances (Jewitt 2009). Neither cometary NEOs nor main-belt comets approach that fraction of activity within their embedded populations, suggesting a closer relationship between the members of the centaur population and comets.

As evidenced from recent cometary missions (cf. Sierks *et al.* 2015; A'Hearn *et al.* 2011), the various surfaces of cometary bodies undergo a significant and rapid surface evolution that manifests as global physical characteristics. Published observations by *Spitzer* have provided a varied picture of the mass-loss rates amongst Centaurs, with either lower, constant mass loss rates ~50 kg/s, as observed for 29P/Schwassmann-Wachmann 1 (Stansberry *et al.* 2004; Schambeau *et al.* 2015; Figure 7a) or the more stochastic but intense mass loss events like the ~$10^3$ kg/s found for the 2006 outburst of Centaur 174P/(60558) Echeclus (Bauer *et al.* 2008; Figure 7b). *Spitzer* thermal observations of active centaur dust affirmed that Centaurs have likely undergone significant surface evolution while in the vicinity of the giant planets.

## IV.  KBOs

Kuiper Belt Objects live beyond Neptune with orbital semimajor axes above 30 au but below about 2000 au, i.e. where the Oort cloud begins (Gladman 2008). *Spitzer* made some of the first observations of significant numbers of KBOs at wavelengths beyond the K band. *Spitzer* photometrically observed thermal heat radiated from ~40 KBOs using MIPS (Rieke et al., 2004). In spite of its small aperture (85 cm), *Spitzer* was so sensitive that it could detect heat radiated from these relatively small (~ 30 - 1000 km), very distant, and cold (40 – 80 K) objects, providing constraints on their reflectivity (albedo) and sizes for the first time (Brucker et al. 2009; Fig. 8). Coupled with supporting optical photometry and the NEATM model (Harris 1998) to derive the albedos and diameters of the targets, *Spitzer* found that the four largest KBOs appear to constitute a distinct class in terms of their albedos (as they have enough mass to trap volatiles gravitationally; Schaller & Brown 2007). *Spitzer* also found that the geometric albedo of KBOs and Centaurs is correlated with their orbital perihelion distance due to increased insolative heating, devolatilization, and mass loss processes for darker objects having smaller perihelia, and that the albedos of KBOs (but not Centaurs) are positively correlated with size. Among the *Spitzer* samples were several binary systems, and the diameters and system masses derived from these provided the first measurements of the bulk density of TNOs (e.g. Stansberry *et al.* 2012; Grundy *et al.* 2015). In some cases, the thermal data were also used to constrain the roughness and thermal inertia of the surfaces, and in the case of Makemake to argue for the presence of extremely low-albedo terrains on an otherwise high albedo object (Lellouch, 2013). Later studies made with the Herschel Space Telescope, with bands at 70, 100 and 160 μm (e.g. Müller *et al.* 2009, 2020; Fornasier et al. 2013, Vilenius *et al.* 2014, 2018), relied heavily on the *Spitzer* 24 μm data to constrain the thermophysical state of TNO surfaces.

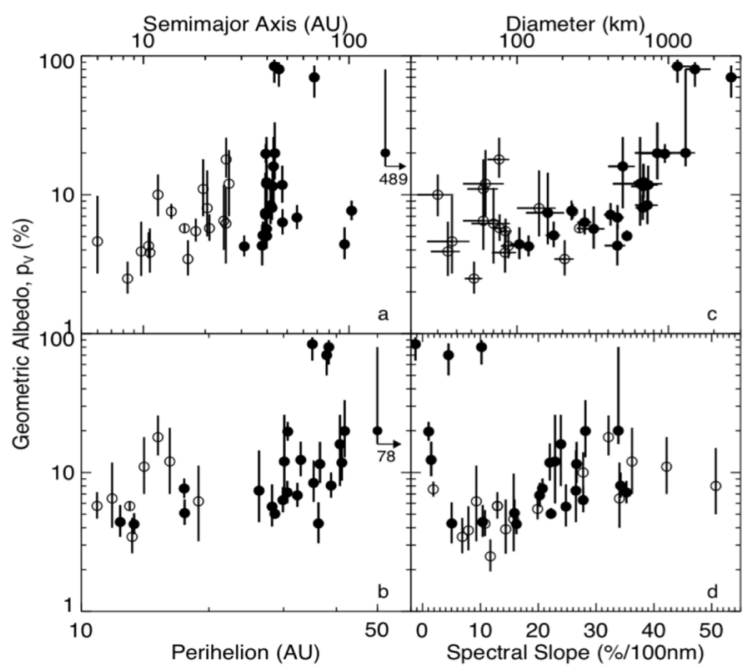

**Figure 8** - Geometric albedo of TNOs based on MIPS thermal observations vs. orbital semimajor axes (top left), perihelion distance (bottom left), object diameter (top right), and spectral slope at visible wavelengths (more positive values correspond to redder surfaces). Filled symbols are for objects with semimajor axes $a > 30$ AU ('TNOs'); open symbols are for those with $a > 30$ AU ('Centaurs'). Pluto is not plotted. The dwarf planets Eris, Makemake and Haumea are the 3 highest-albedo points clustered in all 4 panels. Figure is reproduced from Stansberry et al., 2008.

The combination of visible wavelength and IRAC photometry at 3.6 and 4.5 um provides a powerful method for constraining the surface composition of TNOs (e.g. Emery *et al.* 2007; Dalle Ore *et al.* 2009). Because the absorption features at the IRAC wavelengths are much stronger than those in the near-IR for the same material, TNOs display a much larger range of colors at these wavelengths than they do shortward of 2 μm, and the colors for individual materials are nicely separated. Figure 9 (Fernández-Valenzuela *et al.* 2020) shows a color-color diagram that is diagnostic of the different materials found on the surface of TNOs (e.g., Barucci *et al.* 2020, Young *et al.* 2020). Synthetic reflectance models using a range of grain sizes were constructed for those solid materials, and their convolution with the IRAC bandpasses resulted in colors occupied by the oval shaded regions in Figure 9. Trinary models (i.e., models constructed from combinations of three different materials using different proportions) are represented by the triangle shaded regions. Targets dominated by pure material will share colors with the pure synthetic models, falling within the oval regions, and those with multiple components on their surfaces could have colors intermediate between the pure-component regions. While the IRAC filter bandpasses were too broad to allow definitive detection of specific materials, for those objects with high quality near-IR spectra (only those are shown in the figure), the agreement with the IRAC compositional diagnostics is excellent.

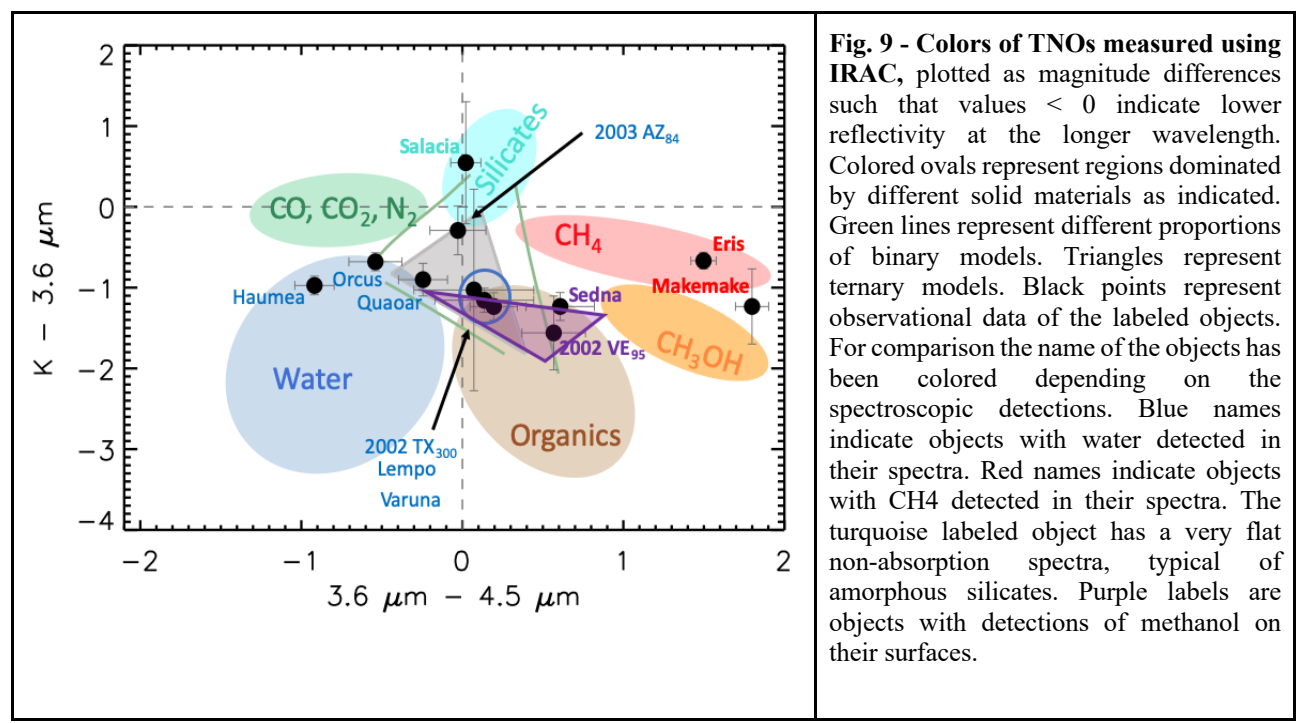

**Fig. 9 - Colors of TNOs measured using IRAC,** plotted as magnitude differences such that values < 0 indicate lower reflectivity at the longer wavelength. Colored ovals represent regions dominated by different solid materials as indicated. Green lines represent different proportions of binary models. Triangles represent ternary models. Black points represent observational data of the labeled objects. For comparison the name of the objects has been colored depending on the spectroscopic detections. Blue names indicate objects with water detected in their spectra. Red names indicate objects with CH4 detected in their spectra. The turquoise labeled object has a very flat non-absorption spectra, typical of amorphous silicates. Purple labels are objects with detections of methanol on their surfaces.

## V.     Conclusions

The contributions of *Spitzer* to planetary science during its 16-year mission were many and varied. Especially when taken together with the subsequent WISE 3.6 – 23 um survey mission, it can be said that "*Spitzer* went narrow and deep while WISE went broad and shallow" on relic bodies produced at the beginning of our Solar System. It is, however, impossible to list all of *Spitzer*'s Solar System observations in this short review; the reader is encouraged to read Part 2 of this paper, as well as the Supplemental On-line Material and to follow up with the cited references. *Spitzer*'s science contributions are also still ongoing and expanding today, due to utilization of the science data archive, and one of the purposes of this review is to highlight potential further studies using the archive. We encourage the reader to contact the authors cited in each subsection to continue this work.

## VI. Acknowledgements

The authors would like to thank first and foremost NASA, JPL, Caltech, and the *Spitzer* project, without which none of the science described above would have been possible. As a NASA mission, local Solar System science measurements could have been downplayed or marginalized, but this was never the case. Instead, the authors experienced *Spitzer* observing schedules built around some of their time critical observations, and large amounts of Legacy science dedicated to their surveys. Project staff were at the same time welcoming & friendly but also highly professional and competent. The science return of NASA efforts like the Deep Impact, STARDUST, and OSIRIS-ReX missions and the ISON and Oumuamua observing campaigns were greatly enhanced by *Spitzer*'s observations. Notable support was provided by many, including L. Armus, S. Carey, C. Grillmair, G. Helou, R. Hurt, V. Meadows, L. Rebull, N. Silberman, G. Squires, T Soifer, L. Storri-Lombardi, and M. Werner. E. Fern'andez-Valenzuela also acknowledges support from the 2017 Preeminent Postdoctoral Program ($P^3$) at UCF.

## **References for Centaurs (30)**

## **References for KBOs (18)**